\documentclass[%
 reprint,
nofootinbib,
 amsmath,amssymb,
 aps,
]{revtex4-1}
\usepackage[utf8]{inputenc}%tildes

\usepackage{graphicx}% Include figure files
\usepackage{dcolumn}% Align table columns on decimal point
\usepackage[scale=0.85]{geometry}
\usepackage{bm}% bold math
\usepackage{bigstrut}
\usepackage{xcolor}
\DeclareUnicodeCharacter{2212}{\textminus}
\usepackage[normalem]{ulem}
\usepackage{float}
\usepackage{hyperref}
\usepackage{physics}
\begin{document}

\preprint{APS/123-QED}
\title{Feasibility to probe the dynamical  scotogenic model at the LHC}

\author{Gustavo Ardila-Tafurth}
\email{ga.ardila10@uniandes.edu.co}

\author{Andrés Flórez}%
\email{ca.florez@uniandes.edu.co}

\author{Cristian Rodríguez}%
\email{c.rodrigez45@uniandes.edu.co} 

\author{Maud Sarazin}
\email{m.sarazin@uniandes.edu.co}
\affiliation{Departamento de F\'isica, Universidad de los Andes, Bogot\'a 111711, Colombia}%

\author{\'Oscar Zapata}%
\email{oalberto.zapata@udea.edu.co}
\affiliation{Instituto de Física, Universidad de Antioquia, Calle 70 \# 52-21,\\
Apartado Aéreo 1226, Medellín, Colombia}%

\newcommand{\oz}[1]{{\color{blue}[OZ: #1]}}
\newcommand{\MS}[1]{{\color{violet}[MS: #1]}}
\newcommand{\ga}[1]{{\color{red}[GA: #1]}}
\date{\today}

\begin{abstract}
    We perform a feasibility study to probe dark matter (DM) production at the LHC within a global $U(1)_L$ scotogenic model. The study is conducted using the Markov Chain Monte Carlo numerical method, considering the viable parameter space of the model allowed by experimental constraints such as neutrino oscillation data, the Higgs to invisible branching fraction, and DM observables. The production of scalar and fermionic DM candidates, predicted by the model, is then studied under the LHC conditions for different luminosity scenarios imposing compressed mass spectra conditions between the lightest fermion and the $\mathbb{Z}_2$ odd scalars. We studied two production mechanisms, Drell-Yan and Vector Boson Fusion. It was found that the Drell-Yan mechanism gives better detection prospects for fermionic DM masses between 100-220~\textrm{GeV} at high luminosity scenarios.
\end{abstract}

\maketitle
\section{Introduction}
\label{Sec:Introduction}
% ======================================================
%Generic itroduction about DM and generation of neutrino masses

The Standard Model (SM) of particle physics, formulated in the 1970s, is one of the most precise theoretical frameworks in physics. It describes  particle interactions through electromagnetic, weak, and strong forces with precision up to 12 significant figures~\cite{ParticleDataGroup:2024cfk}. The SM employs the principle of gauge invariance to explain fundamental interactions between fermions, which are the elementary building blocks of known matter. These interactions are mediated by gauge bosons and are framed within the  $\text{SU}(3)_C\times \text{SU}(2)_L\times \text{U}(1)_Y$ symmetry group. In this model, elementary particles acquire mass through the Higgs mechanism, based on the concept of  electroweak symmetry breaking (EWSB)~\cite{higgspaper, PhysRevLett.13.321, PhysRevLett.13.508, PhysRevLett.13.585}.

Despite being our most accurate theory to date, there are several theoretical questions and experimental conundrums that the SM does not answer. These include the matter-antimatter asymmetry~\cite{Morrissey:2012db, Canetti:2012zc, Davidson:2008bu}, the particle nature of dark matter (DM)~\cite{BertoneHooperSilk2005, Buckley:2017ijx, Boddy:2022knd},  
the origin of neutrino masses~\cite{PhysRevLett.81.1562, SNO:2002tuh}, among others. These limitations suggest that the SM may be just the low-energy limit of a  broader theory. Therefore, a plethora of extended  versions of the SM, referred to as Beyond Standard Model (BSM) scenarios, have been proposed aiming to address these current issues.

A particularly compelling hypothesis addressing the particle nature of DM is the Weakly Interacting Massive Particle (WIMP) scenario. WIMPs feature  masses and interaction strengths at the electroweak scale~\cite{Peebles:1982ff,colddm}, naturally providing the observed relic density through thermal freeze-out~\cite{Steigman:2012nb} in the early universe. Due to their appealing theoretical properties and testability in direct and indirect detection experiments, WIMPs remain among the most extensively studied DM candidates~\cite{Roszkowski:2017nbc,DMwimp}.

Regarding the origin of neutrino masses, experimental observations of neutrino oscillations demonstrate the existence of at least two non-zero mass eigenstates, in clear conflict with the SM prediction of massless neutrinos~\cite{NuFit2020, deSalas:2020pgw}. One of the most prominent neutrino mass generation methods is the so-called Tree-Level Seesaw Mechanism~\cite{Minkowski:1977sc,Yanagida:1980xy, PhysRevLett.56.561, PhysRevD.22.2227, GellMann:1979vob}. This type of mechanism provides an elegant explanation for neutrino masses with minimal additions to the SM particle content. However, high-scale Seesaw scenarios require either extremely heavy new states or very small Yukawa couplings, placing them beyond the reach of current and near-future experimental tests. To reduce the energy scale of new physics and construct experimentally testable models, several alternative strategies have been proposed, including higher-dimensional neutrino mass operators, small lepton-number-violating interactions, and radiative neutrino mass generation mechanisms via loop processes~\cite{Bonnet:2012kz}.

Among the models that generate neutrino masses through loop processes, scotogenic models are particularly appealing, as they unify the origin of neutrino masses and DM, treating them not as separate problems but as interconnected pieces of the same puzzle~\cite{Restrepo2013, Law:2013saa}. In the simplest scotogenic model~\cite{Tao:1996vb,Ma:2006km},  the SM is extended with a $\textrm{SU}(2)_L$ doublet, $\eta$, and three Majorana-like fermion singlets, $N_i$. These new particles  transform as odd states under a $\mathbb{Z}_2$ symmetry, that stabilizes the DM candidates. This model naturally gives rise to a rich DM phenomenology~\cite{Ma:2006km,Klasen:2013jpa,Ibarra:2016dlb}, since the DM candidate can be either a scalar or a fermion. Neutrino masses are then generated radiatively at one-loop level, with the new particle content circulating in the loop, leading to collider and lepton flavor violating signals~\cite{AristizabalSierra:2008cnr,Vicente:2014wga}.
Beyond the simplest realization, other scotogenic models have been explored in several studies~\cite{Toma:2013zsa, Vicente:2014wga, Fraser:2014yha, Baumholzer:2019twf,Restrepo:2015ura}, and their phenomenological implications have been examined in numerous subsequent works~\cite{Molinaro:2014lfa, Longas:2015sxk, Lindner:2016kqk, Rocha-Moran:2016enp, Ahriche:2017iar, Betancur:2017dhy, Bhattacharya:2017sml, Bhattacharya:2018fus, Ahriche:2018ger, konar:2020wvl, Escribano:2020iqq,Sarazin:2021nwo,deNoyers:2024qjz}. 

This paper explores an extension of the scotogenic model, referred to as the \textit{dynamical scotogenic model}~\cite{bonilla2020fermion, U1ScotoValentina, Chun:2023vbh}. The model extends the symmetry group by introducing a global $\textrm{U}(1)_L$ symmetry, which is then spontaneously broken by the nontrivial vacuum expectation value (VEV) of a singlet scalar field, $\sigma$. This breaking induces a mass term for the $N_i$ thanks to their Yukawa interactions with the $\sigma$ scalar~\footnote{Pioneering work in this sector was carried out in~\cite{Mohapatra:1980yp, Mohapatra:1981}, which is considered an extension of the type I tree-level Seesaw Mechanism.}, and delivers a pseudo-Goldstone boson —the Majoron $J$—, whose interactions with leptons give rise to new lepton flavor-violating (LFV) processes.  Additionally, the presence of new fields leads to novel DM interactions in the early universe, enriching the phenomenology with respect to the simplest scotogenic model.   

The possibility of testing the scotogenic model, as well as some of its extensions, at particle colliders has been proposed in several phenomenological studies~\cite{Singh2025TypeIII, Lozano2025Collider, C:2024kds, von_der_Pahlen_2016, Avila:2022jvq}. However, a detailed study to determine the experimental feasibility of probing the dynamical scotogenic model has not yet been conducted. 
This work presents a comprehensive phenomenological study of the collider feasibility  of the dynamical scotogenic model, building upon preliminary analyses performed in Refs.~\cite{bonilla2020fermion, U1ScotoValentina, Chun:2023vbh}. We investigate the detectability of the DM candidates by considering both Drell-Yan (DY) and vector boson fusion (VBF) production mechanisms at the LHC. Our study first determines the viable parameter space, ensuring compatibility with DM, neutrino oscillation data, LFV processes, and Higgs invisible decay constraints through a Markov Chain Monte Carlo (MCMC) analysis. We identify DY processes involving fermionic DM as the most promising detection strategy. Such processes could be probed at the High-Luminosity LHC for DM masses in the range of 100 to 220~\textrm{GeV}.

 The paper is organized as follows. First, Sec.~\ref{Sec:Model} introduces the model, including its particle content, interactions, and mass spectrum. In Sec.~\ref{Sec:Methodological_Setup}, we present the methodological framework used to explore the parameter space of the model, incorporating all relevant experimental constraints to identify the viable regions featuring either a pseudo-scalar or a fermionic DM candidate. 
 Sec.~\ref{Sec:Results} is devoted to the analysis of the impact of direct and indirect detection experiments on the allowed parameter space. In this section, we also  simulate event generation for LHC conditions, obtaining the corresponding production cross-sections as a function of the DM candidate mass. Then, we evaluate the feasibility of probing different benchmark points at the LHC for various luminosity scenarios. Finally, in Section~\ref{Sec:Conclusion} we summarize our findings and present our conclusions.

% ======================================================
\section{The model}%check redaction
\label{Sec:Model}
% ======================================================
The dynamical scotogenic model~\cite{bonilla2020fermion, U1ScotoValentina,Chun:2023vbh} extends the fermion content of the SM by including three additional Majorana fermions ($N_i$, $i = \{1,2,3 \}$). 
The lepton number conservation is ensured by introducing a global $\textrm{U}(1)_L$ symmetry, which is spontaneously broken by the VEV of a new singlet scalar field $\sigma$. 

Besides this singlet, a $\textrm{SU}(2)_L$ scalar doublet field ($\eta$) is also included to couple the SM leptons to the Majorana fermions, thus giving rise to non-zero radiative masses for the SM neutrinos. All the $N_i$ and the extra scalar doublet are considered to be odd under an additional $\mathbb{Z}_2$ symmetry while SM particles and the additional singlet scalar are even. This simple approach stabilizes the lightest $Z_2$-odd particle of the model, naturally becoming a feasible DM candidate. Depending on the specific realization, the DM particle can be either scalar or fermionic, a distinction that shapes the allowed parameter space. The particle content and symmetries of the model are described in Tab.~\ref{Tab:QuantumNumbers}. 

The $\mathbb{Z}_2\times$ U(1)$_L$ invariant scalar potential is given by
\begin{align}
        \mathcal{V}=&\, m_\mathrm{H}^2 \mathrm{H}^\dagger \mathrm{H}+ m_\eta^2\eta^\dagger\eta+m_\sigma^2\sigma^\dagger\sigma +\frac{\lambda_1}{2}(\textrm{H}^\dagger \textrm{H})^2\nonumber\\
        &+\frac{\lambda_2}{2}(\eta^\dagger\eta)^2  +\frac{\lambda_\sigma}{2}(\sigma^\dagger\sigma)^2+\lambda_3(\mathrm{H}^\dagger \mathrm{H})(\eta^\dagger\eta)\nonumber\\
        &+\lambda_3^{\mathrm{H}\sigma}(\mathrm{H}^\dagger \mathrm{H})(\sigma^\dagger\sigma)+\lambda_3^{\eta\sigma}(\eta^\dagger\eta)(\sigma^\dagger\sigma)
        +\lambda_4 (\mathrm{H}^\dagger\eta)(\eta^\dagger \mathrm{H})\nonumber\\
        &+\frac{\lambda_5}{2}[(\mathrm{H}^\dagger \eta)^2+(\eta^\dagger \mathrm{H})^2], 
\end{align}
where the SM Higgs ($\mathrm{H}$) and $\eta$ doublets are parametrized as
\begin{equation}
    \eta ~=~ \begin{pmatrix} \eta^+ \\ \frac{\eta_R + i \eta_I }{\sqrt{2}} \end{pmatrix}\,, \quad \mathrm{H}~=~ \begin{pmatrix} \mathrm{H}^+ \\ \frac{S_\mathrm{H} + i A_\mathrm{H}+v }{\sqrt{2}} \end{pmatrix} \,,
    \label{eq:etaAndHparametrization}
\end{equation}
while for the scalar singlet  
\begin{equation}
    \sigma = \frac{1}{\sqrt{2}}(S_{\sigma} + i A_{\sigma} + v_{\sigma}) \,.
    \label{eq:sigmaParametrization}
\end{equation}

\begin{table}
    \centering
    \begin{tabular}{|c||c|c|c||c|c|c|}
    \hline
            \     & ~$
            \ell_i$~ & ~$\mathrm{e}_{Ri}$~ & ~$N_i$~ & ~$\mathrm{H}$~ & ~$\sigma$~ & ~$\eta$~ \bigstrut\\
            \hline
            \hline
      $\textrm{SU}(2)_L$   &  $\mathbf{2}$ & $\mathbf{1}$ & $\mathbf{1}$ & $\mathbf{2}$ & $\mathbf{1}$ & $\mathbf{2}$ \bigstrut\\
      \hline
        $\textrm{U}(1)_Y$  & $-\frac{1}{2}$ & -1 & 0 & $\frac{1}{2}$ & 0 & $\frac{1}{2}$\bigstrut\\
        \hline
        $\textrm{U}(1)_L$ & 1 & 1 & 1 & 0 & -2 & 0\bigstrut\\
        \hline
            $\mathbb{Z}_2$  & 1 & 1 & -1 & 1 & 1 & -1\bigstrut\\
        \hline 
    \end{tabular}
    \caption{Particle content of the dynamical scotogenic model and their representations under the electroweak gauge groups and global symmetries. }
    \label{Tab:QuantumNumbers}
\end{table}

The charge assignment ensures that only the SM leptons, and not quarks, couple to the dark sector. 
The relevant interaction terms in the Lagrangian density for the new fermionic sector are given by
\begin{equation}
    \mathcal{L} ~\supset~ y_{ij} \bar{\ell_i} \tilde{\eta} N_j + \kappa_{ij} \overline{N_i}^cN_j \sigma + {\rm h.c.} \,; \, i,j=1,2,3.
    \label{Eq:LagrangienF}
\end{equation}
In Eq.~\ref{Eq:LagrangienF} the term $\kappa$ is a $3 \times 3$ matrix, which in our study is considered to be diagonal without loss of generality.

\subsection{Scalar and  fermion spectrum at tree level}
\label{Sec:ScalarMasses}

Because of the $\lambda_3^{\rm H\sigma}$ term in the potential, the \textrm{CP}-even components of the $\mathrm{H}$ and the $\sigma$ scalar fields mix to form two physical eigenstates, $h_1$ and $h_2$, defined as 
\begin{equation}
    \begin{pmatrix} h_1 \\ h_2 \end{pmatrix} ~=~  \begin{pmatrix} {\rm cos}\, \alpha & {\rm sin}\, \alpha \\-{\rm sin}\, \alpha & {\rm cos}\, \alpha \end{pmatrix} \begin{pmatrix} S_\mathrm{H} \\ S_{\sigma} \end{pmatrix} \,.
\label{Eq:Potential}
\end{equation}
From this mixing, we identify  $h_1$  as the SM Higgs boson, with an effective mass of $m_{h_1} \approx 125\,\mathrm{GeV}$~\cite{ATLAS:2015yey}.  From the potential, the tree-level masses of the two physical scalar states are given by\footnote{In our study, all analytical and numerical computations, except for the neutrino mass generation, are performed at tree level.}
\begin{equation}
    m_{h_{1,2}}^2 ~=~ \frac{\lambda_1}{2}v^2+\frac{\lambda_\sigma}{2}v_\sigma^2\mp\frac{1}{2}\Delta\,,
    \label{eq:himasses}
\end{equation}
with $\Delta=\sqrt{(2\lambda_3^{\mathrm{H}\sigma}v v_\sigma)^2+(\lambda_1 v^2-\lambda_\sigma v_\sigma^2)^2}$. 
The mixing angle $\alpha$ can be found in terms of the couplings via the expression 
\begin{align}
       \tan\alpha&=\frac{2\lambda_3^{\mathrm{H}\sigma}v v_\sigma}{\lambda_1v^2-\lambda_\sigma v_\sigma^2- \Delta}\, \label{eq:tanalpha}. %\\[0.65cm]\cos(\alpha) &~=~ \frac{B}{\sqrt{A^2 + B^2}}\,,
\end{align}
This expression is relevant for computing the limit on the branching ratio for the Higgs to invisible decays, as outlined in~Ref.~\cite{PDG2020}: 
\begin{equation}
    \text{cos}(\alpha)^2 \, \text{BR}(h \to \text{invisible}) < 0.11 \,.
    \label{Eq:Htoinv}
\end{equation}
In addition, as the $\textrm{CP}$-odd fields do not mix, $A_{\rm H}$ corresponds  to the longitudinal component of  the SM $\mathrm{Z}$ boson, while $A_{\sigma}$ can be identified as a new massless Goldstone boson (the Majoron). 

Regarding the $Z_2$-odd scalar fields, there are no mass mixing terms; consequently, their masses are 
\begin{equation}
    \begin{split}
        &m_{\eta_R}^2=m_\eta^2 +\frac{1}{2}\lambda_3^{\eta\sigma} v_\sigma^2 +\frac{1}{2}(\lambda_3+\lambda_4+\lambda_5)v^2,\\
        & m_{\eta_I}^2=m_\eta^2 +\frac{1}{2}\lambda_3^{\eta\sigma} v_\sigma^2 +\frac{1}{2}(\lambda_3+\lambda_4-\lambda_5)v^2,\\
        &m_{\eta^\pm}^2=m_\eta^2+\frac{1}{2}(\lambda_3 v^2+\lambda_3^{\eta\sigma}v_\sigma^2).
    \end{split}
    \end{equation}
 
The extra fermions acquire their masses at tree level via the $\kappa$ term, with 
\begin{equation}
    M_{N_{\alpha}} ~=~ \sqrt{2} \, \kappa_{\alpha \alpha} \, v_{\sigma} \,,
    \label{Eq:MXmass}
\end{equation}
where we assume a hierarchical mass spectrum, that is $M_{N_1}<M_{N_2}<M_{N_3}$.

The DM candidate in this model is identified as the lightest neutral $\mathbb{Z}_2$-odd particle. In principle, three fields may fulfill this role: the Majorana fermion $N_1$, the CP-even scalar $\eta_R$, and the CP-odd scalar $\eta_I$. However, for positive values of the $\lambda_5$ coupling, a mass splitting arises that renders $\eta_R$ heavier than $\eta_I$, thereby leaving $N_1$ and $\eta_I$ as the only viable candidates. In this work, we investigate the phenomenology associated with each of these scenarios.

% ======================================================
\subsection{Neutrino masses}
\label{Sec:NeutrinoMasses}
% ======================================================
%
\begin{figure}
    \centering
    \includegraphics[scale=0.5]{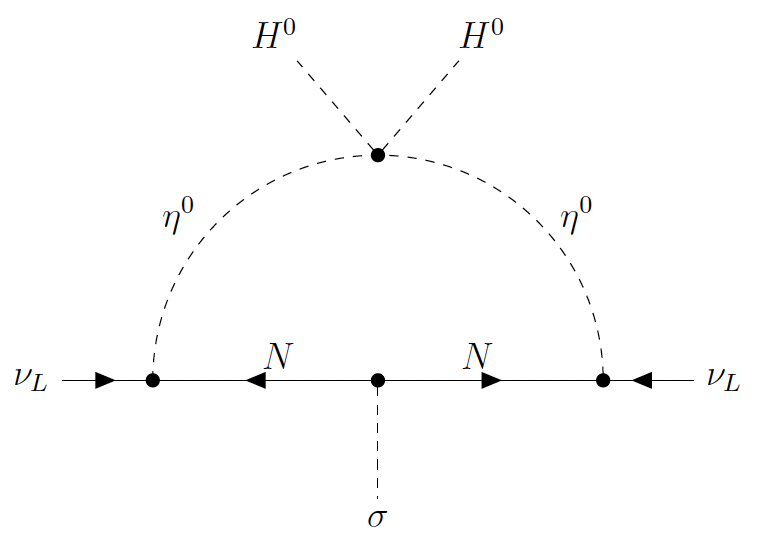}
    \caption{ Representative Feynman diagram illustrating the generation of the neutrino mass matrix element $~(M_{\nu})_{ij}~$ in the physical basis. The $\eta^0$ state represents both the neutral scalar $\eta_R$ and the pseudo-scalar $\eta_I$, while $N$ denotes any of the three new Majorana fermions running in the loop.}
    \label{Fig:neutrinoMasses}
\end{figure}

Neutrino masses are generated radiatively through the $y_{ij}$ interactions in Eq.~\eqref{Eq:LagrangienF}, as illustrated in Fig.~\ref{Fig:neutrinoMasses}. The elements of the $3\times 3$ neutrino mass matrix take the form~\cite{Ma:2006km}
\begin{multline}
    \big( M_{\nu}\big)_{ij} ~=~  \sum_{k=1}^3 \frac{y_{ik} y^{*}_{jk}}{32\pi^2} M_{N_k}\bigg[ \frac{m_{\eta_R}^2}{m_{\eta_R}^2-M_{N_k}^2}\ln \bigg(\frac{m_{\eta_R}^2}{M_{N_k}^2}\bigg) \\- \frac{m_{\eta_I}^2}{m_{\eta_I}^2-M_{N_k}^2}\ln \bigg(\frac{m_{\eta_I}^2}{M_{N_k}^2}\bigg)\bigg] \,.
    \label{Eq:MnuMatrix}
\end{multline}
This mechanism naturally accommodates three non-zero neutrino masses. Assuming a normal mass ordering, $m_1 < m_2 < m_3$, the parameter space must be aligned with current experimental observations for the squared mass differences~\cite{NuFit2020, deSalas:2020pgw}, $\Delta m_{21}^2$ and $\Delta m_{31}^2$. These stringent requirements, however, place tight constraints on the Yukawa couplings, often making it challenging to identify compatible solutions. To overcome this, we adopt the Casas-Ibarra parametrization~\cite{CasasIbarra2001}, where  the Yukawa couplings are expressed  in terms of masses and neutrino mixing angles as
\begin{equation}
    y = \sqrt{\Lambda^{-1}} O \sqrt{\hat{M_\nu}} U^{\dagger}_{\text{PMNS}}\,.
    \label{Eq:CouplingMatrixCasasIbarra}
\end{equation}
Here, $y$ is the Yukawa coupling matrix, $O$ is an arbitrary $3\times 3$ orthogonal matrix,  $U_{\text{PMNS}}$ is the PMNS mixing matrix, and $\Lambda~=~$Diag$(\Lambda_i)$ is a diagonal matrix encoding the loop functions, with
\begin{multline}
    \Lambda_i=\frac{M_{N_i}}{32\pi^2}\bigg[\frac{m_{\eta_R}^2}{m_{\eta_R}^2-M_{N_i}^2}\ln\bigg(\frac{m_{\eta_R}^2}{M_{N_i}^2}\bigg)\\-\frac{m_{\eta_I}^2}{m_{\eta_I}^2-M_{N_i}^2}\ln\bigg(\frac{m_{\eta_I}^2}{M_{N_i}^2}\bigg)\bigg]\,. 
\end{multline}

% ======================================================
\section{Methodological setup}
\label{Sec:Methodological_Setup}
% ======================================================

The first step of our study is to analyze the behavior of the relic abundance of the given DM candidate as a function of its mass. The model is implemented and its particle spectrum is generated using the software {\tt SARAH} (version  4.15.1)~\cite{SARAH2010,SARAH2011,SARAH2013,SARAH2014}, which is then used by the {\tt SPheno} package (version  4.0.5)~\cite{SPheno2003,SPheno2012,spheno4}, at leading order, to compute low-energy observables and rates of LFV processes, and to estimate the Higgs to invisible branching ratio. The relic abundance, spin-independent DM-nucleon scattering cross-section and the average annihilation cross-section times velocity for indirect detection are calculated using {\tt micrOMEGAs} (version 5.3.41)~\cite{Belanger:2018ccd}. 
All relevant experimental constraints from Tab.~\ref{Tab:constraints} are incorporated into our analysis via an MCMC~\cite{Markov1971} numerical code. Furthermore, using {\tt FeynRules} (version 2.3)~\cite{ALLOUL20142250}, we generate a Universal Feynman Output (UFO)~\cite{Degrande:2011ua} module for {\tt MadGraph5\_aMC\@NLO} (version 3.1.0)~\cite{Alwall:2011uj}. The {\tt xslha} python package (version 0.2.2)~\cite{Staub2019_xSLHA} is then used to transfer the {\tt SPheno} output into the \texttt{MadGraph} parameter cards, in order to calculate the production cross-sections for collider analysis at the LHC.

\subsection{Free parameters}
\label{Sec:Parameters}

The scalar potential contains 12 parameters: four dimensionful and eight dimensionless. The two minimization conditions, together with the two scalar masses $m_{h_{1,2}}$, allow us to fix  four of these parameters. We choose them to be $m_{\rm H}^2, m_\sigma^2, \lambda_1$ and $\lambda_\sigma$. Consequently, the remaining free scalar parameters are 
\begin{align}
    \{\lambda_2, \lambda_3,\lambda_4,\lambda_5,\lambda_3^{\eta \sigma},\lambda_3^{\rm H\sigma}, m_\eta^2,v_\sigma\}.
\end{align}
For the new fermion sector, we consider the $\kappa_{\beta\beta}$ couplings as free parameters. Finally, in order to implement the Casas–Ibarra parametrization, $m_{\nu_1}$ is varied freely along with the arbitrary angles in the orthogonal matrix $\mathcal{O}$.
All in all, the free parameters are varied within the ranges displayed in Tab.~\ref{Tab:parameter_ranges_Setup}.

\begin{table}[]
    \centering
    \begin{tabular}{|c|c|}
        \hline
         \textbf{Observable} & \textbf{Constraint}
         \bigstrut\\
         \hline\hline
        $m_{h_1}$ & 125.25 $\pm$ 3.0 \textrm{GeV}
        \bigstrut\\
        \hline
        $\text{BR}(\mu^-$ $\to$ $\mathrm{e}^- \gamma$ ) & $< 4.2 \times 10^{-13}$
        \bigstrut\\
        \hline
        $\text{BR}(\tau^-$ $\to$ $\mathrm{e}^- \gamma$ ) & $< 3.3 \times 10^{-8}$
        \bigstrut\\
        \hline
        $\text{BR}(\tau^-$ $\to$ $\mu^- \gamma$ ) & $< 4.2 \times 10^{-8}$
        \bigstrut\\
        \hline
        $\text{BR}(\mu^-$ $\to$ $\mathrm{e}^- \mathrm{e}^+ \mathrm{e}^-$ ) & $< 1.0 \times 10^{-12}$
        \bigstrut\\
        \hline
        $\text{BR}(\tau^-$ $\to$ $\mathrm{e}^- \mathrm{e}^+ \mathrm{e}^-$ ) & $< 2.7 \times 10^{-8}$
        \bigstrut\\
        \hline
        $\text{BR}(\tau^-$ $\to$ $\mu^- \mu^+ \mu^-$ ) & $< 2.1 \times 10^{-8}$
        \bigstrut\\
        \hline
        $\text{BR}(\tau^-$ $\to$ $\mathrm{e}^- \mu^+ \mu^-$ ) & $< 2.7 \times 10^{-8}$
        \bigstrut\\
        \hline
        $\text{BR}(\tau^-$ $\to$ $ \mu^- \mathrm{e}^+ \mathrm{e}^-$ ) & $< 1.8 \times 10^{-8}$
        \bigstrut\\
        \hline
        $\text{BR}(\tau^-$ $\to$ $ \mu^- \mathrm{e}^+ \mu^-$ ) & $< 1.7 \times 10^{-8}$
        \bigstrut\\
        \hline
        $\text{BR}(\tau^-$ $\to$ $ \mu^+ \mathrm{e}^- \mathrm{e}^-$ ) & $< 1.5 \times 10^{-8}$
        \bigstrut\\
        \hline
            $\text{BR}(\tau^-$ $\to$ $  \mathrm{e}^- \pi $ ) & $< 8.0 \times 10^{-8}$
        \bigstrut\\
        \hline
        $\text{BR}(\tau^-$ $\to$ $ \mathrm{e}^- \eta$ ) & $< 9.2 \times 10^{-8}$
        \bigstrut\\
        \hline
        $\text{BR}(\tau^-$ $\to$ $ \mathrm{e}^- \eta'$ ) & $< 1.6 \times 10^{-7}$
        \bigstrut\\
        \hline
        $\text{BR}(\tau^-$ $\to$ $  \mu^- \pi $ ) & $< 1.1 \times 10^{-7}$
        \bigstrut\\
        \hline
        $\text{BR}(\tau^-$ $\to$ $ \mu^- \eta$ ) & $< 6.5 \times 10^{-8}$
        \bigstrut\\
        \hline
        $\text{BR}(\tau^-$ $\to$ $ \mu^- \eta'$ ) & $< 1.3 \times 10^{-7}$
        \bigstrut\\
        \hline
        $\text{CR}_{\mu \to \mathrm{e}}$(Ti) & $< 4.3 \times 10^{-12}$
        \bigstrut\\
        \hline
        $\text{CR}_{\mu \to \mathrm{e}}$(\textrm{Pb}) & $< 4.3 \times 10^{-11}$
        \bigstrut\\
        \hline
        $\text{CR}_{\mu \to \mathrm{e}}$(Au) & $< 7.0 \times 10^{-13}$
        \bigstrut\\
        \hline
        $\text{BR}(\mathrm{Z}^0$ $\to$ $ \mathrm{e}^\pm \mu^\mp  $ ) & $< 7.5 \times 10^{-7}$
        \bigstrut\\
        \hline
        $\text{BR}(Z^0$ $\to$ $ \mathrm{e}^\pm \tau^\mp $ ) & $< 5.0\times 10^{-6}$
        \bigstrut\\
        \hline
        $\text{BR}(\mathrm{Z}^0$ $\to$ $ \mu^\pm \tau^\mp$ ) & $< 6.5 \times 10^{-6}$
        \bigstrut\\
        \hline
        $\text{BR}(h \, \to \,$ \text{invisible} ) & $< 0.11 \times  \text{cos}(\alpha)^{-2} $
        \bigstrut%\\
        %\hline
        %$\abs{\sin(\alpha)}$  & $ < 0.3$
        \bigstrut\\
        \hline
    \end{tabular}
    \caption{Experimental constraints applied in the numerical analysis, with the majority provided as upper limits at the $90\%$ confidence level.}
    \label{Tab:constraints}
\end{table}

\begin{table}[]
\centering
\begin{tabular}{|c|c|c|}
\hline
\textbf{Parameter} & \textbf{Range} \bigstrut\\
\hline
\(\lambda_2\) & \([10^{-2}, 1]\) \bigstrut\\
\(\lambda_3\) & \([10^{-2}, 1]\) \bigstrut\\
\(\lambda_4\) & \([10^{-2}, 1]\) \bigstrut\\
\(\lambda_5\) & \([10^{-10}, 1]\) \bigstrut\\
\(\lambda_3^{\eta \sigma}\) & \([10^{-2}, 1]\) \bigstrut\\
\(\lambda_3^{\rm H \sigma}\) & \([10^{-7}, 10^{-5}]\) \bigstrut\\
\hline
\(m_{\eta}^2\) & \([9 \times 10^4, 2.5 \times 10^7]\) \bigstrut\\
%\hline
\(v_\sigma\) & \([10^3, 10^4]\) \bigstrut\\
\hline
%$\sin \alpha$ & $[-0.3,0.3]$\bigstrut\\
%\hline
\(\kappa_{11}\) & \([10^{-2}, 1]\) \bigstrut\\
\(\kappa_{22}\) & \([10^{-2}, 1]\) \bigstrut\\
\(\kappa_{33}\) & \([10^{-2}, 1]\) \bigstrut\\
\hline
\(m_{\nu_1}\) & \([10^{-32}, 10^{-12}]\) \bigstrut\\
\hline
\end{tabular}
\caption{Values of the selected input parameters for the numerical scan, with all mass parameters expressed in units~of~[\textrm{GeV}].}
\label{Tab:parameter_ranges_Setup}
\end{table}

% ======================================================
\subsection{The numerical method}
\label{Sec:NumericalMethod}

The analysis of the model follows Refs.~\cite{Sarazin:2021nwo, Alvarez:2023dzz, deNoyers:2024qjz}, using a {\tt Python} based Metropolis-Hastings~\cite{Metropolis1953} algorithm, allowing us to efficiently scan, through an iterative process, the viable parameter space of the model. 

The individual likelihood associated with each constraint is computed as
\begin{equation}
   {\rm ln }(\mathcal{L}_i) ~=~ \frac{(\theta_{obs}^i - \theta^i_{exp})^2}{2 \sigma_{i}^2}\,,
   \label{Eq:IndividualLikelihood}
\end{equation}
where $\theta_{obs}^i$ represents the value of a specific observable estimated at a point $``i"$, $\theta^i_{exp}$ is the associated experimental result, and $\sigma_i$ is the corresponding uncertainty,  based on the list in Tab.~\ref{Tab:constraints}, which includes experimental bounds on the SM Higgs boson mass, the Higgs to invisible branching ratio, and constraints from LFV measurements. The final likelihood is then estimated from the product of all the individual likelihoods, assuming that the constraints are not correlated. The estimation is performed using a step function smeared as a single-sided Gaussian with a width of $10\%$ relative to the experimental data, using Eq.~\eqref{Eq:IndividualLikelihood}. In cases where the observed value $\theta_{obs}^i$ is below the upper limit, the likelihood is equal to one.   

Because of the mixing in the scalar sector,  the SM Higgs  signal strength gets modified by a factor of $\cos^2\alpha$. Hence,  it follows that the condition $|\sin\alpha|\lesssim 0.3$ must be fulfilled for $m_{h_2}> m_{h_1}$~\cite{Falkowski:2015iwa,Arcadi:2019lka,Ferber:2023iso}, from the most recent LHC results~\cite{ATLAS:2022vkf,CMS:2022dwd}. 
In addition, direct searches for additional Higgs bosons give stronger bounds on $\sin\alpha$ than the Higgs signal strength for $150\,{\rm GeV}< m_{h_2} \lesssim 700$ \textrm{GeV}~\cite{Ferber:2023iso}. Models that do not satisfy these collider bounds are excluded from our sample, independently of their likelihood.

In addition, the presence of a massless Majoron in the spectrum gives rise to both astrophysical and cosmological constraints. From the astrophysical perspective, Majorons can be copiously produced in the cores of stellar objects such as red giants and white dwarfs, where their emission provides an additional channel for energy loss.
To remain consistent with well-established astrophysical observations, the rate of Majoron emission must therefore be sufficiently suppressed, which requires the Majoron–lepton couplings to be very small.
In the present analysis, however, these constraints are significantly relaxed, since the Majoron–lepton interactions arise only at the one-loop level and are thus loop-suppressed. The assessment of the impact of these constraints are beyond the scope of this study and will be left as a secondary analysis.  Additionally, as this study considers a massless Majoron, its presence could potentially affect the effective number of neutrino species, $\rm N_{eff}$, prior to recombination. As noted in Refs.~\cite{U1ScotoValentina, Chun:2023vbh}, these effects could be significantly reduced if $\lambda_3^{\rm H\sigma}\lesssim 10^{-5}$.

After selecting the points that satisfy all the conditions outlined in Tab.~\ref{Tab:constraints}, we compute the relic DM density. We then exclude points that exceed the experimental upper bound, specifically applying the $3\sigma$ constraint $\Omega h^2 < 0.12$ from the PLANCK collaboration~\cite{Planck2018}. The parameter space is further constrained by a suite of experimental bounds arisen from  direct~\cite{LZ2023,Bo2025PandaX4T,Aprile2024XENONnT,Aalbers2016} and indirect~\cite{FermiLAT2015,Atwood2009} searches for DM;
bounds on the oblique parameters $S$, $T$, and $U$~\cite{Peskin1992Estimation,ParticleDataGroup:2024cfk} ($S= -0.01 \pm 0.10, \quad T= 0.03 \pm 0.12, \quad U = 0.02 \pm 0.11$); and exclusion bounds from the LHC~\cite{Aad2020, SUSYCMS2019}. We analyze the impact of these results on the model's parameter space by selecting the points that fall within the corresponding allowed regions.  

Our numerical MCMC analysis focuses on a \textit{compressed mass spectrum} scenario. This choice is motivated by phenomenological studies~\cite{VBFPhysRevD.87.035029, VBFPhysRevD.90.095022, VBFPhysRevLett.111.061801, VBFPhysRevD.91.055025,natalia2021, CMS:2019zmn, SUSYCMS2019, CMS:2016ucr,CMS:2015jsu}. Specifically, we require $\Delta m = |m_{\eta_R} - M_{N_1}| <  30\,\mathrm{GeV}$ (50 \textrm{GeV}). This requirement has an important implications on the relic DM abundance below the \textrm{TeV} scale, as well as for the density of viable solutions identified in the numerical scan\footnote{ 
Compressed mass scenarios have been extensively explored in the context of supersymmetric models at the LHC, where they correspond to regions of parameter space that remain unconstrained by experimental searches~\cite{Sirunyan2019, Cardona2022, PhysRevD.107.075010, qureshi2024probing, PhysRevD.107.115026, Agin:2024yfs, Dutta:2023jbz, das:2024xle, PhysRevD.94.073007, Avila:2018sja, Ballabene:2022fms, Rossini:2019wrc}.}.
The subset of parameter space consistent with all experimental limits is subsequently employed to perform a feasibility study of the detection prospects for each DM candidate at the LHC. The results of this comprehensive analysis are presented in Sec.~\ref{Sec:Results}. 
% ======================================================
\subsection{Collider Analysis Setup}
\label{Sec:ColliderStudy}
% ======================================================
  
\begin{figure}
    \centering
    \includegraphics[width=0.48\linewidth]{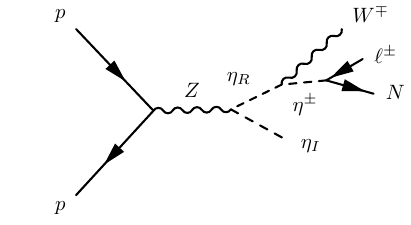}
    \includegraphics[width=0.48\linewidth]{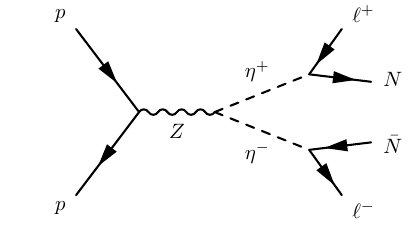}
    \caption{Feynman diagrams associated with the production of DM via DY processes at the LHC. Left: sample topology for the production of the scalar DM candidate, $\eta_I$, along with a Majorana fermion and SM particles as final states. Right: sample topology for the production of the fermionic DM candidate, $N_1$, along with two SM leptons.}
    \label{fig:Feynman_Diagrams_Full_DY}
\end{figure}

A direct implication of a compressed mass spectrum is the low momentum of SM leptons arising from the $\eta^{0(\pm)} \to N_i + \nu_j (\ell^{\pm})$ decays. The reconstruction, identification, and analysis of final states with low momentum leptons can be challenging at the experimental level.  
At the LHC, production from $\mathrm{q} \bar{\mathrm{q}}$ fusion can give rise to signals with potential to be probed through DY processes such as  $\mathrm{p+p} \to \eta_{I} +\eta_R \to \eta_I +\eta^{\pm} +\mathrm{W}^{\mp} \to \eta_I +\ell^\pm+ N_i+\mathrm{W}^\mp$ and $\mathrm{p+p} \to \eta^{\pm}+ \eta^{\mp}\to N_{i} +\ell^{\pm}+ N_{i}+\ell^{\mp}$, where $\ell^{\pm}$ represents a SM charged lepton. Fig.~\ref{fig:Feynman_Diagrams_Full_DY} shows representative Feynman diagrams for the DY production of scalar (left) and fermionic (right) DM signal processes at the LHC.  

Purely neutral final states can be probed through VBF signatures. Experimentally, VBF processes are characterized by final states with charged leptons or missing transverse energy ($E_{\mathrm{T}}^{miss}$) in the detector's central region, accompanied by two energetic jets with a large pseudorapidity gap, $\Delta \eta(j_1, j_2) > 3.8$, and large dijet invariant mass,
\begin{equation}
    m_{jj} \approx \sqrt{2p_{T}^{j_1}p_{T}^{j_2} \cosh(\Delta \eta(j_1, j_2)) } > 500 \, \mathrm{GeV}.
\end{equation}

\begin{figure}
    \centering
    \includegraphics[width=0.36\linewidth]{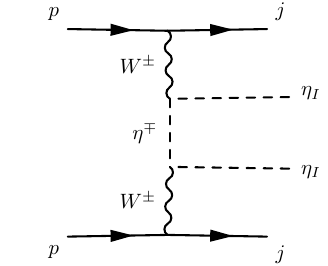}
    \includegraphics[width=0.54\linewidth]{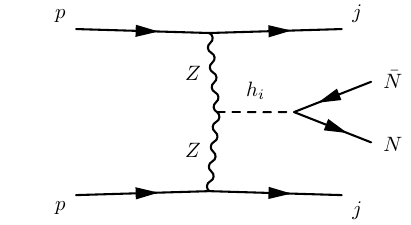}
    \caption{Feynman diagrams illustrating the production of DM via VBF processes. The scalar DM channel is depicted on the left, while the fermionic channel is shown on the right.}
    \label{fig:Feynman_Diagrams_Full_VBF}
\end{figure}

This type of topology has been extensively explored in phenomenological studies as a powerful tool for searching for new physics at the LHC~\cite{VBFPhysRevD.87.035029, VBFPhysRevD.90.095022, VBFPhysRevLett.111.061801, VBFPhysRevD.91.055025, natalia2021}, and it has also been employed in experimental searches~\cite{SUSYCMS2019, CMS:2016ucr, CMS:2015jsu}. Moreover, VBF has proven valuable in Higgs boson to invisible studies and in probing challenging regions of parameter space, such as those associated with compressed mass spectra in SUSY searches~\cite{20121, 201230, SUSYCMS2019, Aad2020}. In the context of scotogenic models, VBF processes are particularly relevant due to the presence of Higgs portal interactions and the coupling of the inert scalar doublet $\eta$ to electroweak gauge bosons. These features may allow VBF to access regions that remain inaccessible via DY production mechanisms. However, no dedicated study has yet been conducted in this direction. In the case of the dynamical scotogenic model, VBF production of DM particles can proceed either via resonant Higgs production or through $\textrm{W}$-mediated processes. Figure~\ref{fig:Feynman_Diagrams_Full_VBF} shows representative Feynman diagrams for the VBF production of scalar (left) and fermionic (right) DM signal processes at the LHC.

With the objective of producing DM particles under a parameter space compatible with both the observed DM relic density and different experimental constraints enumerated in Tab.~\ref{Tab:constraints}, we transfer the values of the couplings and masses obtained from the scans to a set of files compatible with {\tt MadGraph}. For this part of the analysis, we consider scenarios with $\Delta m < (30, 50)$~\textrm{GeV} and for the specific values $m_{h_2} = (246, 500)$~\textrm{GeV}.

% ======================================================
\section{Phenomenological Analysis}
\label{Sec:Results}
% ======================================================

% ======================================================
\subsection{DM Relic abundance} 
\label{Subsec:DM}
% ======================================================
Following the methodology outlined in the previous Section, we first analyze the DM relic density as a function of the DM candidate mass, while incorporating all the constraints listed in Tab.~\ref{Tab:constraints}. 

\begin{figure}
    \centering
    \includegraphics[width=\linewidth]{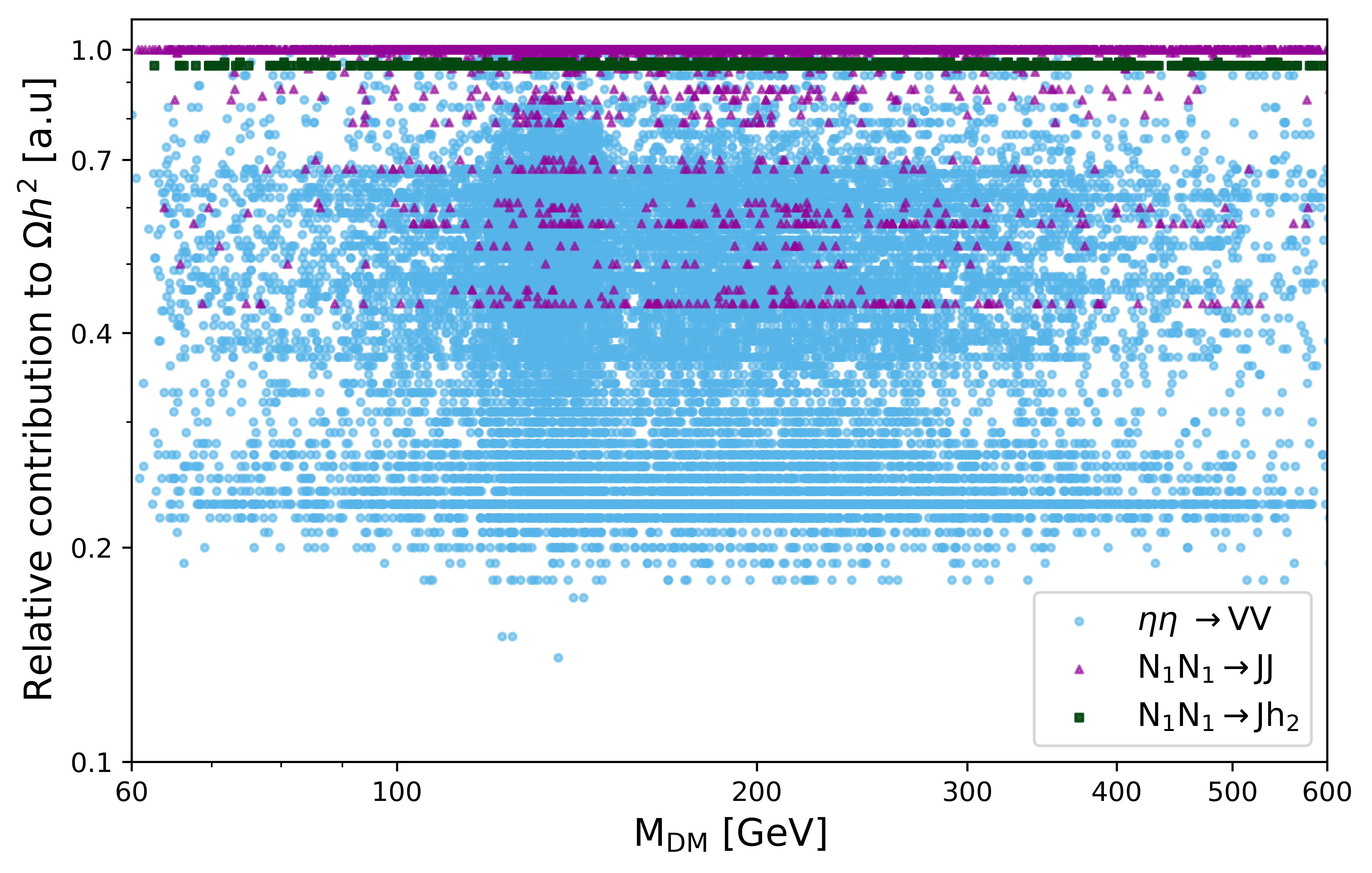}
    \caption{Relative contribution of each processes to the relic density projected on the DM mass  when $N_1$ is the DM candidate and for $m_{h_2}=246\,\textrm{GeV}$ and $\Delta m < 30~\textrm{GeV}$.}
    \label{Fig:N1_DM_processes}
\end{figure}

\begin{figure}
    \centering
    \includegraphics[width=\linewidth]{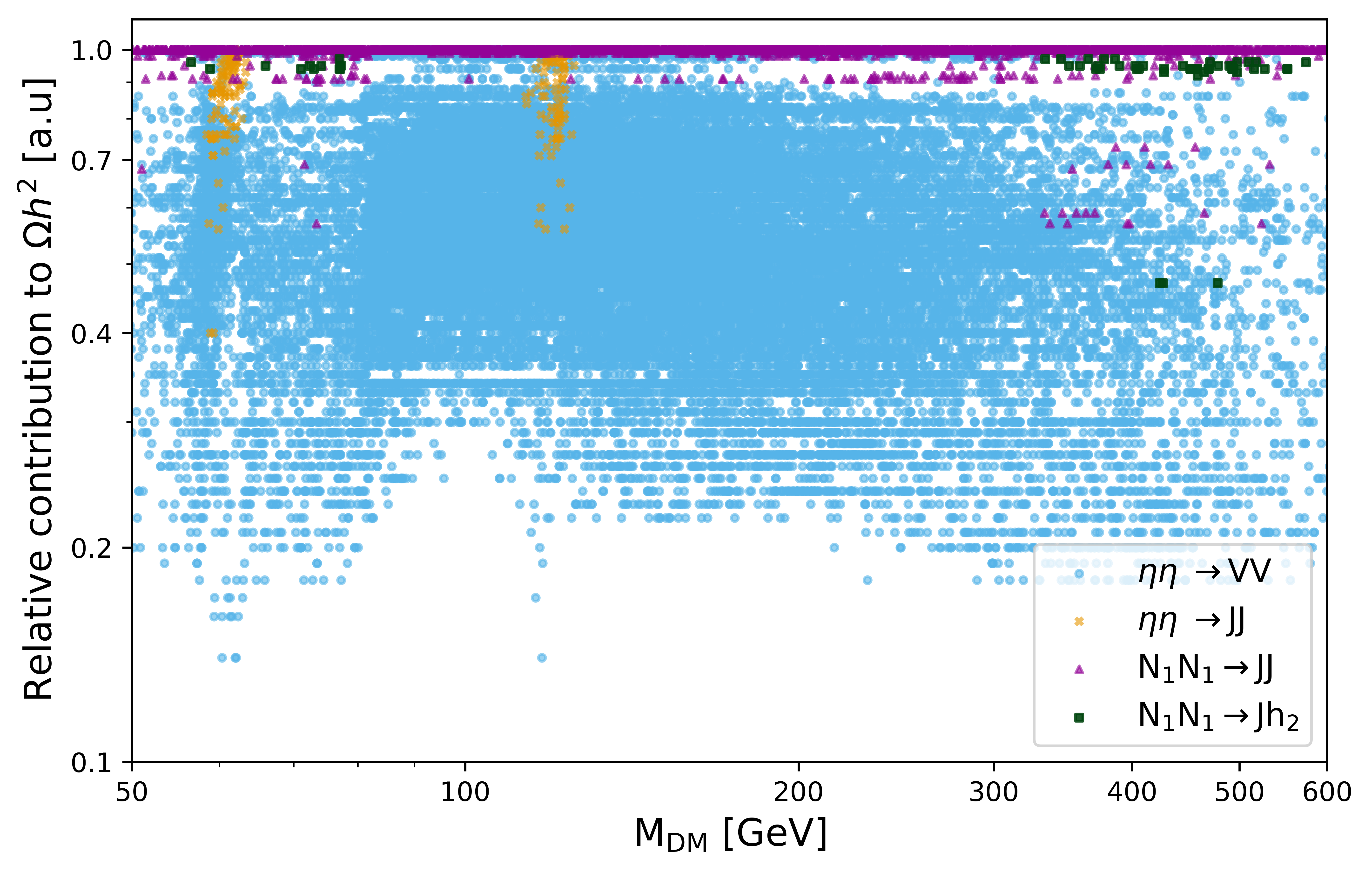}
    \caption{Relative contribution of each processes to the relic density projected on the DM mass  when $\eta_I$ is the DM candidate and for $m_{h_2}=246\,\textrm{GeV}$ and $\Delta m < 30~\textrm{GeV}$. 
    }
    \label{Fig:EtI_DM_processes}
\end{figure}

By imposing a compressed mass spectrum, the small mass splitting enhances the efficiency of coannihilation channels, making them play a significant role in determining the DM relic density. This behavior is illustrated in Figs.~\ref{Fig:N1_DM_processes} and~\ref{Fig:EtI_DM_processes}, where we have fixed $m_{h_2}=246~\textrm{GeV}$ and $\Delta m < 30~\textrm{GeV}$ (similar results are found for $m_{h_2}=500$ GeV).

For the case where $N_1$ is the DM candidate, the dominant annihilation channel is $N_1 N_1 \to JJ$, while the $N_1 N_1\to Jh_2$ process is subdominant. In addition, annihilation processes of the inert doublet $\eta$ into weak gauge bosons contribute significantly to the total DM annihilation rate, becoming in some cases the most dominant ones across the entire mass range of interest.

In the scenario where the pseudo-scalar $\eta_I$ is the DM candidate and no mass splitting is imposed, the standard inert-doublet behavior is recovered: for $m_{\eta_I} \lesssim 500~\textrm{GeV}$ the relic density is suppressed due to efficient gauge-mediated annihilations. When a compressed mass spectrum is enforced, however, $N_1$-induced processes may also become relevant. As illustrated in Fig.~\ref{Fig:EtI_DM_processes}, fermionic channels involving $N_1$, such as $N_1 N_1 \to JJ$ and $Jh_2$, can contribute significantly to the relic density. Furthermore, the $\eta\eta\to JJ$ process provides large contributions to the abundance in the resonance regions corresponding to $\frac{m_{h_1}}{2}$ and $\frac{m_{h_2}}{2}$, in contrast to the fermion DM scenario where the $h_1$ funnel region is suppressed by the small values of $\lambda_3^{\rm H\sigma}$ coupling.  
 
% ======================================================
\subsection{DM Direct and indirect detection}
\label{Subsec:ID}
% ======================================================

In both fermion and scalar scenarios, the direct detection of DM occurs through its elastic scattering off nuclei, mediated by the scalars $h_1$ and $h_2$ at tree level. Hence, sizable detection rates can arise in both scenarios. 
Since parameter-space points with a relic abundance below the observed value are also considered viable in this analysis, a rescaling factor $\xi$ is introduced, defined as
\begin{align}
    \xi=\frac{\Omega_{DM}^n}{\Omega_{\text{PLANCK}}} \,,
    \label{Eq:xi}
\end{align}
where $\Omega_{\text{PLANCK}}h^2= 0.12$ is the central experimental value for the DM relic density reported by the PLANCK collaboration~\cite{Planck2018}, while $\Omega_{\rm DM} h^2$ is the relic density calculated for each point.

\begin{figure}
    \centering
    \includegraphics[width=\linewidth, height=0.25\textheight]{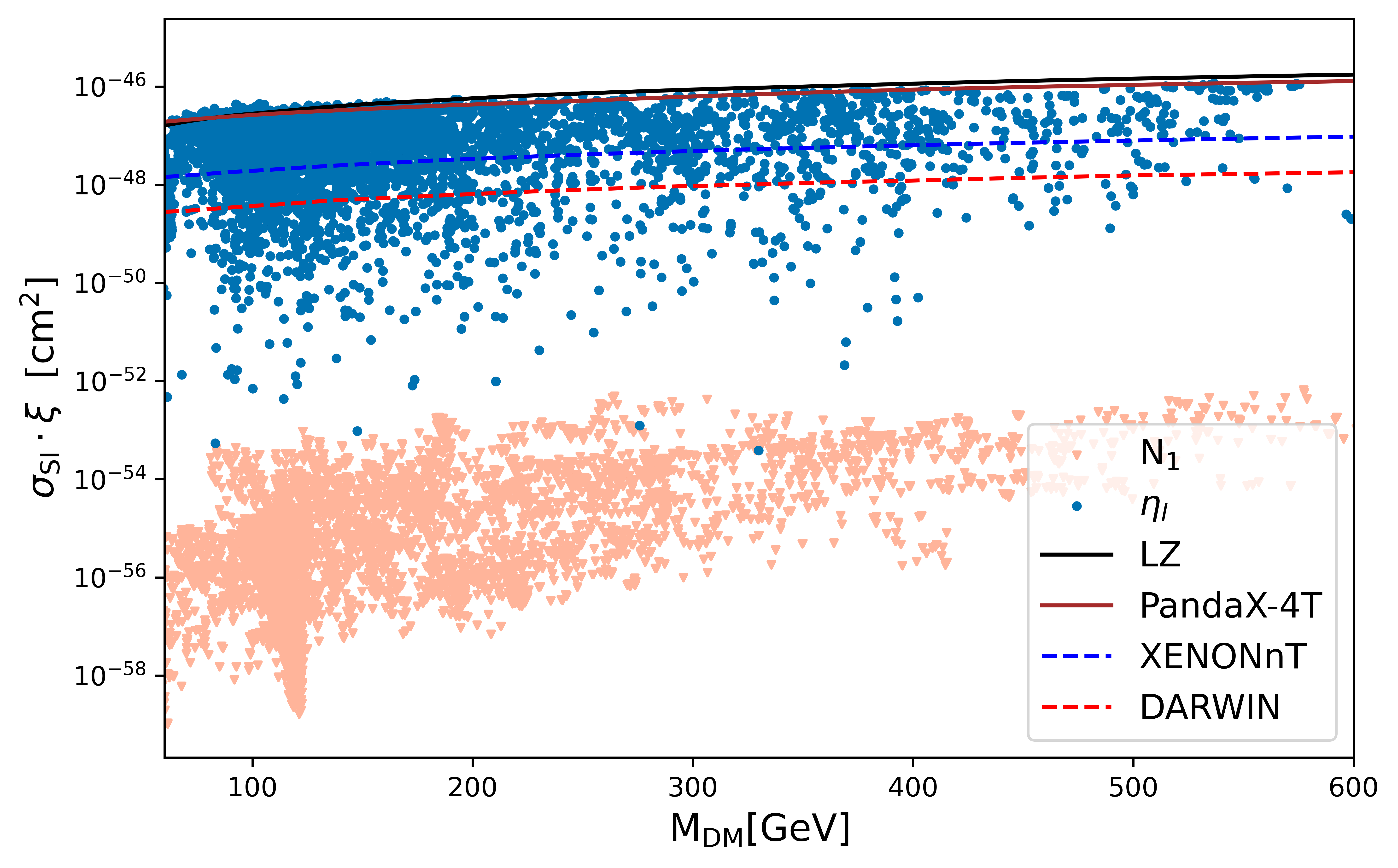}\\
    \caption{ Rescaled spin-independent cross-section as a function of DM mass, for the two different DM candidates, $N_1$ (pink triangles) and $\eta_I$ (purple circles). The solid lines represent the current upper limits reported by LZ and PandaX-4T experiments while dashed lines represent the expected sensitivity from XENONnT and DARWIN. The points shown in the plot are consistent with the DM relic density constraint (within or below the current bound)  and are within the compressed mass sprectrum for $\Delta m<30~ \rm GeV$ and $\Delta m<50~\rm GeV$. 
    }
    \label{fig:DD_no_fixed_vev}
\end{figure}

Figure~\ref{fig:DD_no_fixed_vev} displays the rescaled spin-independent DM-nucleon scattering cross-section, $\xi\sigma_{SI}$, as a function of the DM mass, for a fixed value of  $m_{h_2}=246~\rm GeV$ noting that varying the value has a mild impact on the DM phenomenology. The solid lines correspond to the upper limits reported by the LZ~\cite{LZ2023} and PandaX-4T~\cite{Bo2025PandaX4T} collaborations, while the dashed lines indicate the projected sensitivities for XENONnT and DARWIN experiments~\cite{Aprile2024XENONnT,Aalbers2016}. Here, it is shown that the $N_1$-nucleon scattering cross-sections are significantly suppressed to values below \( 10^{-53}\,\mathrm{cm}^2 \). This suppression arises because the interaction proceeds exclusively through $t$-channel exchange of either \( h_1 \) or \( h_2 \). The corresponding vertex function depends explicitly on \( \lambda_3^{H\sigma} \), which is assumed to satisfy $\lambda_3^{H\sigma}<10^{-5}$. As a consequence, the resulting scattering cross sections are driven into the neutrino floor region.
In constrast, when taking into account the projected sensitivities from future experiments such as XENONnT and DARWIN~\cite{Aprile2024XENONnT,Aalbers2016}, large portions of the parameter space corresponding to the $\eta_I$ DM candidate can be probed for masses ranging from 100 to 1000~\textrm{GeV}. 

\begin{figure}
    \centering
    \includegraphics[width=\linewidth, height=0.25\textheight]{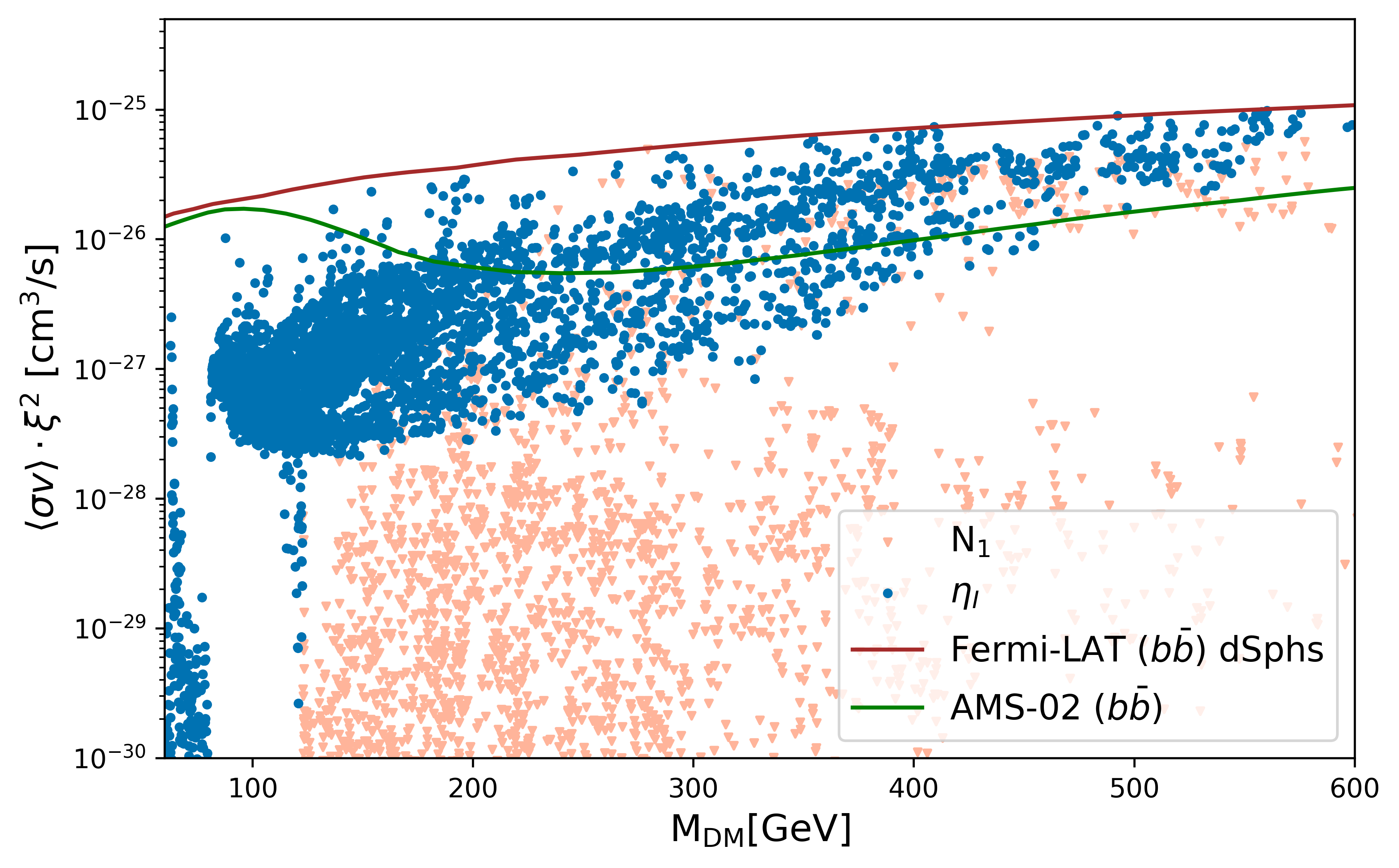}\\
    
    \caption{Rescaled DM annihilation rates 
    %Velocity-averaged total annihilation cross-section
    as a function of the DM mass, for the two DM candidates, $N_1$ (pink triangles) and $\eta_I$ (purple circles), considering $m_{h_2}=246~\rm GeV$ and both $\Delta m<30~ \rm GeV$ and $\Delta m<50~\rm GeV$. The solid lines represent the current limits imposed by the Fermi-LAT gamma ray data~\cite{FermiLAT2015} and by the AMS-02 spectrometer~\cite{Reinert_2018}. The shown points are consistent with the relic density and direct detection upper bounds.}
   
    \label{fig:ID_no_fixed_vev}
\end{figure}

In addition to the direct detection, we also analyze the implications for the indirect detection of DM. 
The most relevant DM annihilation channels are  \(h_1 h_1, h_2 h_2, \mathrm{W}^\pm \mathrm{W}^\mp \). The corresponding rescaled rates are contrasted against the upper limits obtained from the Fermi-LAT data  (in the \( \textrm{b}\bar{\textrm{b}} \) channel) using dSphs~\cite{FermiLAT2015}  assuming dominant $\rm b\bar{b}$ production, as shown in Fig.~\ref{fig:ID_no_fixed_vev}.  Furthermore, the latest AMS-02 spectrometer results~\cite{Reinert_2018}, which include the combined \( \rm \bar{p} \) and B/C ratios under the assumption of dominant \( \rm b\bar{b} \) production, have also been incorporated. However, they were not employed as a constraint owing to the strong model dependence of the associated propagation. Consequently, current indirect detection data do not impose any meaningful restrictions on the allowed parameter space of the model.

%============================================================
\subsection{LHC Signatures}
\subsubsection{Scalar DM scenario}
\label{Subsec:DY}
%============================================================
To estimate the potential to probe the production of pseudo-scalar states $\eta_I$ at colliders via DY mechanism, we calculate the cross-section values corresponding to a  significance of $1.69\,\sigma$. This threshold defines our expected exclusion at 90\% confidence level and is derived from the condition
\begin{align}
    \frac{S}{\sqrt{S+B+\delta_{\text{Sys}}^2}} = 1.69 \,,
    \label{eq:significance}
\end{align}
where $S$ is the expected number of signal events, $B$ the total expected background, and the term $\delta_{\text{Sys}}$ corresponds to systematic uncertainties. Both $S$ and $B$ are proportional to the integrated luminosity $\mathcal{L}$ and their respective production cross-sections. This simplified single-bin counting experiment provides a conservative estimate, as a full binned likelihood analysis exploiting the kinematic distributions and correlations would yield improved sensitivity\footnote{However, as commented in~\cite{Baker:2019sli}, this conservative single-bin approach comes at the cost of reduced sensitivity, since extending the analysis to multiple bins can significantly strengthen the exclusion limits. Nevertheless, this approach is sufficient to establish the discovery potential and guide the phenomenological analysis.}~\cite{Segura:2024srj}.

The estimated production cross-section values for the $\mathrm{pp} \to \eta_{R} \eta_I \to \mathrm{W^{\pm}} \eta^{\pm} \eta_{I} \to  \mathrm{W^{\pm}} N_{i} \ell \eta_{I}$ process are obtained with \texttt{MadGraph5\_aMC}, considering $\mathrm{pp}$ collisions at $\sqrt{s} = 13.6 \, \mathrm{TeV}$ for scenarios with $m_{h_2}=246~\rm GeV$, after including all the experimental constraints from Tab.~\ref{Tab:constraints} and a $\Delta m < 30 \, \mathrm{GeV}$ upper threshold. All points considered in the scan are consistent with the DM relic density, according to the estimates from PLANCK or values that fall below this threshold, as well as the current limits on DM-Nucleon scattering from PandaX-4T and the upper bound on  $\expval{\sigma v}$ from Fermi-LAT.

This analysis considers an optimistic scenario where background events constitute only $20\%$ of the total expected yield. We include a flat $20\%$ systematic uncertainty on the background rate and a $10\%$ detector efficiency factor. The estimation of the observable signal region is performed for three LHC luminosity scenarios~\cite{Henderson:2021qrr}, $\mathcal{L} = \{137, 300, 3000\}~\textrm{fb}^{-1}$, which would be sensitive to cross-section values in the $[10^{-5},10^{-3}]~\rm{pb}$ range. However, our calculated cross-sections for the viable parameter space fall in the $[10^{-18},10^{-10}]~\rm{pb}$ range, remaining several orders of magnitude below the required sensitivity. Consequently, we see that $\eta_I$ production via DY processes is unlikely to be observable at the LHC in the foreseeable future. 

As discussed in Sec.~\ref{Sec:ColliderStudy}, the purely neutral final state can be probed via signatures from VBF processes at colliders. Specifically, we focus on producing signatures involving two pseudo-scalar particles and two jets while explicitly excluding contributions from QCD processes and considering the processes illustrated in the left panel of Fig.~\ref{fig:Feynman_Diagrams_Full_VBF}. Similar final states have been studied in Ref.~\cite{SUSYCMS2019} within the context of SUSY models. We extrapolate their results to our target mass ranges to estimate the detection feasibility for luminosity scenarios of $300 \textrm{ fb}^{-1}$ and $3000\textrm{ fb}^{-1}$, which imply sensitivity within the cross-section range $[10,10^{2}]~\rm pb$. However, we find that the production cross-sections fall in the $[10^{-8}, 10^{-1}] ~\rm pb$ range, just below the estimated limits.

%============================================================ 
\subsubsection{Fermionic DM scenario}
%============================================================

In fermionic DM production via DY mechanisms, the visible decay chain considered is $\eta^\pm\to N_1+ \ell$, as depicted in Fig.~\ref{fig:Feynman_Diagrams_Full_DY}. This process generates two vertices that arise from the decay of each $\eta^{\pm}$, which introduces vertex functions proportional only to the Yukawa coupling $y_{ij}$. This contrasts with the scalar sector, where there is an additional dependence on $\lambda_i$ arising from the $\eta_R$ decay. Consequently, the cross-section takes values from $10^{-2}$~\textrm{pb} down to $10^{-12}$~\textrm{pb},  higher than those in the scalar sector.

\begin{figure}[]
    \centering
    \includegraphics[width=\linewidth]{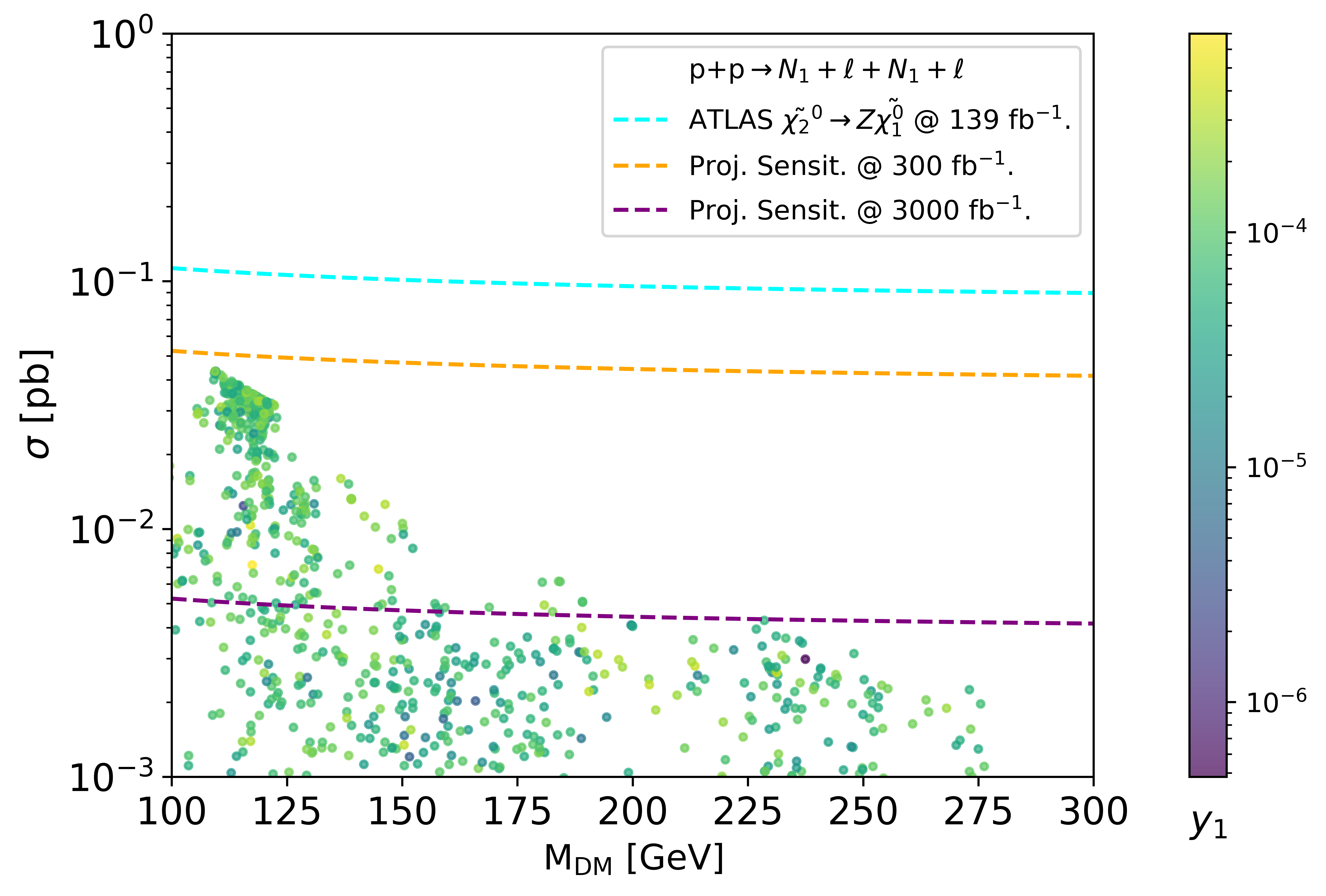}
    \caption{Behavior of the $p+p\to N_i+\ell+N_i +\ell$ cross-section as a function of the fermionic DM mass, for $\Delta m < 30$ \textrm{GeV} and $m_{h_2}=246~\rm GeV$ . Dashed lines denote the experimental limits from ATLAS SUSY studies in compressed mass spectra models~\cite{Aad2020} (Cyan), as well as expected sensitivities at $\mathcal{L}= 300$ \textrm{fb}$^{-1}$ (Orange) and $\mathcal{L}= 3000$ fb$^{-1}$ (Purple). The color-map shows the different values for the $y_1=\sqrt{\abs{y_{N_1 e}}^2 +\abs{y_{N_1\mu}}^2+ \abs{y_{N_1\tau}}^2}$  Yukawa coupling. 
    }
    \label{fig:xsfull_fermion}
\end{figure}

Fig.~\ref{fig:xsfull_fermion} illustrates the behavior of the cross-section as a function of the DM candidate in the fermionic sector. The requirement on $\Delta m < 30$ \textrm{GeV} yields expected final states that consist of soft leptons and invisible particles, similar to those studied in ATLAS SUSY analyzes for compressed mass spectra in Ref.~\cite{Aad2020}. Therefore, we perform a recast of the ATLAS results, assuming the same experimental performance holds at higher luminosities. Our simulations are compared with the observed limits from Ref.~\cite{Aad2020} at $\mathcal{L}=139~\textrm{fb}^{-1}$, and we extrapolate the expected sensitivities to $\mathcal{L}=300~\textrm{fb}^{-1}$ and $\mathcal{L}=3000~\textrm{fb}^{-1}$ under this assumption.

 In contrast to the scalar sector, we observe that a portion is above the expected limits associated with $\mathcal{L}=300$ $\textrm{fb} ^{-1}$ and $\mathcal{L}=3000$ $\textrm{fb} ^{-1}$. These results suggest that fermionic DM could be probed at the High Luminosity LHC for mass ranges between 100 and 220~\textrm{GeV}, in searches targeting high $E_{\mathrm{T}}^{miss}$ in association with  low-momentum charged leptons.

For the purely neutral production of fermionic DM via VBF processes illustrated in the right panel of Fig.~\ref{fig:Feynman_Diagrams_Full_VBF}, we follow the same strategy used in the scalar sector, extrapolating the cross-section values obtained in Ref.~\cite{SUSYCMS2019} to our target mass ranges in order to estimate the detection feasibility for luminosity scenarios of $300~\textrm{fb}^{-1}$ and $3000~\textrm{fb}^{-1}$. In this analysis, the extrapolated values fall in the same range than in the scalar VBF sector ($[10,10^{2}]~\rm pb$) whereas the obtained production cross-sections take values in the $[10^{-8}, 10^{-3}]~\rm pb$ range, well below the aforementioned limits.  The conjunction of DY and VBF results imply that the former mechanism could become a potential probe for fermionic DM, while VBF remains unsuitable for both DM particles even at high luminosity values at the LHC.
%============================================================
\subsection{Scalar DM at FCC-hh}
\label{Subsec:VBF}
%============================================================

The gap between the experimental sensitivity and viable production cross sections could be closed at the Future Circular Collider (FCC) ~\cite{FCC:2018vvp}. This facility promises center-of-mass collision energies approaching $\sqrt{s} \sim 100 \textrm{ TeV}$, which would significantly enhance production rates, combined with ultra-high luminosity $\mathcal L\sim 25\,\textrm{ab}^{-1}$ that could extend the  probed parameter space.

Since the cross-section values obtained in the VBF analysis for the scalar sector lie close to the High-Luminosity LHC (HL-LHC) sensitivity prospects, we recalculated them at a center-of-mass energy of $\sqrt{s} = 100\ \mathrm{TeV}$, corresponding to the proposed center-of-mass energy for the hadronic ideation of FCC (FCC-hh)~\cite{FCC:2018vvp}. These calculations employ the \texttt{NNPDF40\_nlo\_as\_01190} Parton Distribution Functions (PDFs), which offer improved precision in the determination of proton structure~\cite{Ball2022}. Furthermore, we extrapolate the CMS limits and projected sensitivities from Ref.~\cite{SUSYCMS2019} to luminosities of $7\ \textrm{ab}^{-1}$, $15\ \textrm{ab}^{-1}$, and $25\ \textrm{ab}^{-1}$, values within the range expected from a 20-25 year operation of the FCC-hh~\cite{Zimmermann:2016puu}.

\begin{figure}
    \centering
    \includegraphics[width=\linewidth]{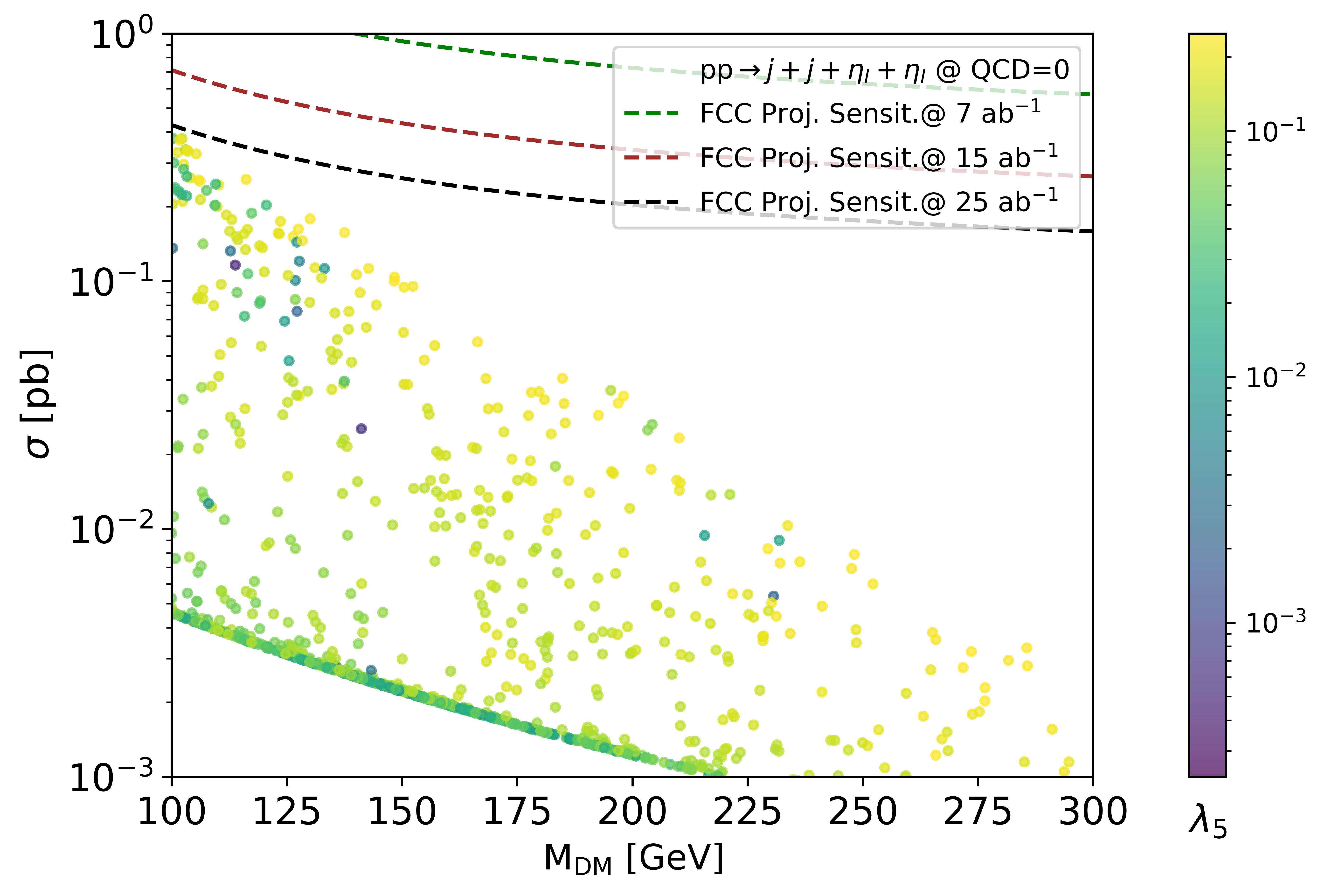}
    \caption{Behavior of the $p+p\to j+j+\eta_I+\eta_I$ cross-section as a function of the DM mass for $\Delta m <50\textrm{ GeV}$. The green, brown and black dashed lines are extrapolations of the SUSY VBF results from the CMS collaboration \cite{SUSYCMS2019} considering FCC luminosity scenarios of 7, 15 and 25~$\textrm{ab}^{-1}$ respectively. 
}
    \label{fig:FCC_vbf}
\end{figure}

Figure~\ref{fig:FCC_vbf} shows the VBF production cross-section as a function of the scalar DM mass, using the aforementioned $\sqrt{s} = 100\,\mathrm{TeV}$ for the FCC-hh. Compared to the LHC results, the cross-sections are notably larger due to phase space enhancement from the higher center-of-mass energy. However, the predicted cross sections fall slightly below the expected exclusion limit at $\mathcal{L}=25~\rm ab^{-1}$. Therefore, moderate improvements in integrated luminosity or analysis sensitivity could render these processes detectable for masses in the 150–200\,\textrm{GeV} range. It is worth noting that such mass values are relatively low for FCC standards, which may exacerbate background and pileup levels and reduce the efficiency of analyzes in this regime. While the FCC-hh will improve the accessible cross-section compared to current experiments, the expected reach remains insufficient to firmly establish the detectability of DM for the mass range under consideration through VBF processes.

 % ======================================================
\section{Summary}
\label{Sec:Conclusion}
% ======================================================

We have conducted a DM and collider feasibility phenomenological study of the dynamical scotogenic model. 

This framework extends the SM by a $\mathbb{Z}_2$ even singlet scalar $\sigma$, a $\mathbb{Z}_2$ odd scalar doublet $\eta$, three Majorana fermions which are also odd under $\mathbb{Z}_2$, and a Goldstone boson (the Majoron) resulting from the $\textrm{U}(1)_L$ symmetry breaking. In this setup, neutrino masses are generated radiatively at one-loop level through interactions involving the new $\mathbb{Z}_2$-odd states. The model naturally accommodates two viable DM candidates: the lightest Majorana fermion $N_1$ and the pseudo-scalar $\eta_I$.

To assess the detectability of these DM candidates, we investigated both Drell–Yan (DY) and vector boson fusion (VBF) production mechanisms at hadron colliders. The analysis was carried out over the viable parameter space identified through a Markov Chain Monte Carlo (MCMC) scan, incorporating all relevant experimental constraints, including the observed DM relic abundance, direct detection limits, collider bounds, and constraints from lepton-flavor-violating processes.  

In the DY processes, we find that the scalar DM scenario consistently yields production cross sections below the projected experimental sensitivities for integrated luminosities of $\mathcal{L}=\{137, 300, 3000\}$ $\textrm{fb}^{-1}$. In contrast, the fermionic DM scenario exhibits cross sections that may be accessible at the High-Luminosity LHC, particularly in final states characterized with by large missing transverse energy and soft leptons, for DM masses between 100 and 220 \textrm{GeV}. These sensitivity estimates are obtained by extrapolating existing ATLAS and CMS analyses performed in supersymmetric frameworks with similar final-state topologies. 

For VBF production, we find that neither DM scenario leads to observable signals at the LHC. The predicted cross sections for scalar DM lie well below the expected sensitivity for luminosities between 300~$\textrm{fb}^{-1}$ and 3000~$\textrm{fb}^{-1}$ and DM masses between 100~\textrm{GeV} to 1~\textrm{TeV}. 
 
Likewise, the cross-section values obtained for the  fermionic scenario in the VBF channel remain far below the sensitivity estimates inferred from current ATLAS and CMS searches, even at the highest luminosities considered. 
These results stand in clear contrast to the DY case, where a limited but phenomenologically interesting region of parameter space remains accessible for fermionic DM.

All in all, our findings indicate that Drell–Yan production constitutes the most promising collider probe of fermionic DM in the dynamical scotogenic model, while VBF production is ineffective for testing either DM candidate, even at high-luminosity collider stages such as the HL-LHC or future facilities like the FCC.
\acknowledgments
The authors express their sincere gratitude to Avelino Vicente and Valentina De Romeri for their valuable discussions and insights regarding the model phenomenology and its implementation. A.F., C.R.,  M.S., and G.A.T. thank the constant and enduring financial support received for this project from the Faculty of Science at Universidad de Los Andes (Bogot\'a, Colombia) through the projects INV-2023-178-2999, INV-2023-175-2957, and INV-2024-199-3203. O.Z. acknowledges partial support from Sostenibilidad-UdeA, the UdeA/CODI Grants 2022-52380 and  2024-76476, and the Ministerio de Ciencias Grant CD 82315 CT ICETEX 2021-1080.

\bibliography{paper.bib}

\end{document}